\documentclass[aip,jcp,reprint,twocolumn,nofootinbib,floatfix,amsmath,amssymb,groupedaddress]{revtex4-1}
\usepackage{graphicx}
\usepackage{color}

\makeatletter
\renewcommand{\@biblabel}[1]{(#1)}
\makeatother

\usepackage[english]{babel}
\usepackage[utf8]{inputenc}
\usepackage{ae}
\usepackage{latexsym}
\usepackage{subfigure}
\usepackage{dcolumn}
\usepackage{bm}

\newcommand{\degree}{\ensuremath{^\circ}}
\newcommand{\rholo}[1]{\rho_{\rm {#1}}}
\newcommand{\rhored}[1]{\rho^{\star}_{\rm {#1}}}
\newcommand{\surft}[1]{\gamma_{\rm {#1}}}
\newcommand{\eps}[1]{\epsilon_{\rm {#1}}}
\newcommand{\rsfig}[1]{
  \begin{center}
    \vspace{0.3cm}
    \includegraphics*[width=8.5cm]{{#1}}
  \end{center}
}

\newcommand{\comment}[1]%
{\textsf{\textcolor{red}{#1}}}

\begin{document}

\title{Statics of polymer droplets on deformable surfaces}

\author{F.~Léonforte}
\author{M.~Müller}
\affiliation{
  Institut für Theoretische Physik, Georg-August-Universität,\\
  Friedrich-Hund-Platz 1, 37077 Göttingen, Germany}

\begin{abstract}
	The equilibrium properties of polymer droplets on a soft deformable surface are investigated by molecular dynamics simulations of a bead-spring model. The surface consists of a polymer brush with irreversibly end-tethered linear homopolymer chains onto a flat solid substrate. We tune the softness of the surface by varying the grafting density. Droplets are comprised of bead-spring polymers of various chain lengths. First, both systems, brush and polymer liquid, are studied independently in order to determine their static and dynamic properties. In particular, using a numerical implementation of an AFM experiment, we measure the shear modulus of the brush surface and compare the results to theoretical predictions. Then, we study the wetting behavior of polymer droplets with different contact angles and on substrates that differ in softness. Density profiles reveal, under certain conditions, the formation of a wetting ridge beneath the three-phase contact line. Cap-shaped droplets and cylindrical droplets are also compared to estimate the effect of the line tension with respect to the droplet size. Finally, the results of the simulations are compared to a phenomenological free-energy calculation that accounts for the surface tensions and the compliance of the soft substrate. Depending on the surface/drop compatibility, surface softness and drop size, a transition between two regimes is observed: from one where the drop surface energy balances the adhesion with the surface, which is the classical Young-Dupré wetting regime, to another one where a coupling occurs between adhesion, droplet and surface elastic energies.
\end{abstract}

\maketitle

\section{Introduction}
\label{sec:intro}

The molecular structure and dynamics at the boundary between a liquid and a confining surface dictate the interactions of a liquid with the surface and the friction of a fluid that flows past it \cite{Bocquet07,Stone,Leonforte11}. They determine the equilibrium shape of drops and the microscopic dissipation mechanisms give rise to slippage, control the (de)-wetting kinetics of thin coating layers \cite{PJG,BW91,Reiter92,Karim,Green03,Fetzer06,Fetzer05,Baumchen09,Rev1} and set the time scale, on which droplets spread on a wettable surface. Since early works of Young \cite{Young1805} and Navier \cite{Navier1823}, the effect of molecular interaction and friction at the surface are described within a continuum theory via boundary conditions, e.g. Young's equation and Navier's slip condition, featuring the surface and interfacial tensions, the hydrodynamic interface position, and the slip length as coarse-grained parameters that transfer the microscopic information to the continuum description.

Which coarse-grained parameters describe the microscopic behavior depends on the specific system. Often, idealized hard planar solid surfaces are considered. Such ideal surfaces, however, are rarely encountered in practice, and the softness of a surface is a natural departure from the ideality of surfaces. These deformable surfaces are expected to differ in their static wetting behavior and the softness also influences the flow of liquids past it.

For instance, when a droplet is deposited on a surface, and the spreading process has been completed, an equilibrium contact angle $\theta_Y$ is reached. Classically, this spreading process should essentially depend on the capillary properties of the liquid and the surface, and also on the liquid viscosity. The equilibrium contact angle for a non-deformable surface, is given by the parallel balance of surface-tension forces at the contact line \cite{Young1805}. For soft elastomeric surfaces, however, a local deformation will occur near the triple line, which is dictated a balance between the component of the surface tension of the liquid perpendicular to the surface and the ability of the soft surface to deform. This interplay gives rise to a lifting-up of the contact line, which has been shown to largely contribute to the wetting or dewetting on soft elastomeric surfaces \cite{Shanahan94,Carre95,Carre96,Kumar04,D1,D2}, as well as to the visco-elastic braking of a driven droplet rolling over a deformable surface \cite{Carre01}.

Molecular dynamics (MD) simulations provide a well suited framework to investigate the physical quantities pertinent to the wetting of  liquids on soft deformable solid surfaces. In the following, we study a polymer brush as a prototypical example of a soft deformable surface, where flexible polymers are tethered with one end to a flat, impenetrable and hard substrate. The ability of the soft surface to deform is controlled by the grafting density of tethered chains. The softness of the surface is, in a sense, universal because it stems from the balance between the loss of conformational entropy as the tethered macromolecules stretch away from the grafting substrate and their excluded volume interactions. Both, in our simulations as well as in experiments, it can be tuned by altering the degree of polymerization of the brush molecules or their grafting density without changing the non-bonded interactions.  This soft surface is in contact with a polymer liquid. Varying the incompatibility between the brush and the polymer liquid we can independently tune the softness and the contact angle that a liquid drops forms on the brush.  

Our work is organized as follows: In sec.~\ref{sec:model} we describe the model and provide details of the simulations. In the Sec.~\ref{subsec:droplet}, the equilibrium properties of droplets properties are studied, and the deformability of the soft surface is characterized in the Sec.~\ref{subsec:brush}. The observation of the equilibrium wetting of cylindrical droplets in contact with different deformable surfaces are discussed in the Sec.~\ref{subsec:cylindrical}. Then, the effects of the line tension are investigated in the  Sec.~\ref{subsec:spherical}, in which spherical, cap-shaped and cylindrical droplets in contact with the same soft surface are considered. Finally, we summarize our conclusions and give a brief outlook.

\section{Model and Simulation aspects}
\label{sec:model}

MD simulations are performed using a well established polymer coarse-grained model \cite{Kremer90}. Each polymer is represented by a chain of $N$ beads of mass $m=1$ connected by springs to form a linear, flexible chain. Beads, that are not neighbors along a chain molecule, interact with a Lennard-Jones (LJ) potential:

\begin{equation}\label{LJpot}
  \mathrm{U^{\alpha\beta}_{\rm LJ}(r)} = 4\epsilon_{\alpha\beta}\Bigl[\left(\frac{\sigma}{r}\right)^{12} - \left(\frac{\sigma}{r}\right)^6\Bigr]\;,
\end{equation}

\noindent for $r < r_{\rm cut} = 2.5\sigma$, while for $r \geq r_c$, the potential is cut-off and shifted such that it is continuous at $r_{\rm cut}$. The potential comprises a steep repulsion at short distances, which mimics the excluded volume of the beads, and a longer-ranged attraction, which makes the polymer condense into a dense liquid that coexists with its vapor of vanishingly low density (poor solvent condition). Indices $(\alpha,\beta)$ stand for the different types of pairs, $dd$, $bb$, and $bd$ for droplet $(d)$ or brush $(b)$. All energies are measured in units of $\epsilon$. Adjacent beads along the chains are coupled through an anharmonic finite extensible nonlinear elastic potential (FENE): 

\begin{equation}\label{FENEpot}
  \mathrm{U_{\rm FENE}(r)} = -0.5 k R^2_0 \ln{\biggl[1 - \left(\frac{r}{R_0}\right)^2\biggr]}\;,
\end{equation}

\noindent where model parameters are identical to those given in Ref.~\cite{Kremer90}, namely $k=30\epsilon/\sigma^2$ and $R_0=1.5\sigma$, chosen in such that unphysical bond crossings and chain breaking are eliminated. The bonded interaction, Eq.~\eqref{FENEpot}, is applied in conjunction with a purely repulsive LJ potential, i.e.~Eq.~\eqref{LJpot} with $r_{\rm cut}=2^{1/6}\sigma$. All quantities are expressed in terms of molecular diameter $\sigma = 1$, LJ energy $\epsilon = 1$, and characteristic time $\tau=\sqrt{m\sigma^2/\epsilon}$.

The classical equation of motion are integrated via the velocity-Verlet algorithm with a time step of $\Delta t=5\times10^{-3}\tau$. Temperature is kept constant at $k_B T=1.2\epsilon$ using a dissipative particle dynamics (DPD) thermostat, that also accounts for hydrodynamic interaction due to the local conservation of momentum \cite{DPD1,DPD2}. The thermostat adds to the total conservative force, that arise from \eqref{LJpot} and \eqref{FENEpot}, a dissipative force $\textbf{F}^D_i$ on each monomer $i$, and a random force, $\textbf{F}^R_i$. Both forces are applied in a pairwise manner, such that the sum of thermostatting forces acting on a particle pair vanishes. Let $\Gamma$ be the friction constant, the dissipative force is given by:

\begin{equation}\label{Fdiss}
	\textbf{F}^D_i=-\Gamma\sum_{j\neq i}\omega^D(r_{ij})\left(\hat{\textbf{r}}_{ij}.	\textbf{v}_{ij}\right)\hat{\textbf{r}}_{ij}\;,
\end{equation}
	
\noindent where $\hat{\textbf{r}}_{ij}=(\textbf{r}_i-\textbf{r}_j)/r_{ij}$ and $\textbf{v}_{ij}=\textbf{v}_i-\textbf{v}_j$. We choose the weight functions:

\begin{equation}\label{Wfunction}
	\omega^D(r_{ij}) = \left\{ \begin{array}{ll}
        \left(1 - r_{ij}/r_c\right)^2 &\mbox{$\,r_{ij} < r_c$}\\
        0 &\mbox{$\,r_{ij} \geq r_c$}
        \end{array}
        \right.\;,
\end{equation}

\noindent with $r_c$ identical to the one used in Eq.\eqref{LJpot}. The random force is given by:

\begin{equation}\label{Frandom}
\textbf{F}^R_i = \xi\sum_{j\neq i}\omega^R(r_{ij})\theta_{ij}\hat{\textbf{r}}_{ij}\;,
\end{equation}

\noindent where $\theta_{ij}$ is a random variable with zero mean, unit variance, second moment $\langle\theta_{ij}(t)\theta_{kl}(t')\rangle = \left(\delta_{ij}\delta_{jl}+\delta_{il}\delta_{jk}\right)\delta(t-t')$, and $\theta_{ij}=\theta_{ji}$. The weight functions $\omega^R(r_{ij})$ satisfy the fluctuation-dissipation theorem, $\left[\omega^R\right]^2=\omega^D$. Friction $\Gamma$ and noise strength $\xi$ define the temperature via $\xi^2=2k_BT\Gamma$. We choose $\Gamma=0.5\tau^{-1}$ in all our simulations.

\begin{table}
	\begin{tabular}{|l||c|c|c|}
	\hline
	N & $10$ & $100$ & $200$ \\
	\hline\hline
	$R_e$ [$\sigma$] & $3.57$ & $12.28$ & $15.2$ \\
	\hline  
	$\eta_d$ [$\sigma^2/\sqrt{m\epsilon}$] & $6.4 \pm 0.1$ & $73 \pm 4$ & $120 \pm 6$\\ 
	\hline  
	$6D$ [$\sigma^2/\tau$]$\times 10^3$ & $65$ & $2.4$ & $0.57$ \\ 
	\hline  
	$\tau_p$ [$\tau$] & $33.1$ & $9.8\,10^3$ & $2.9\,10^4$ \\ 
	\hline
	$\gamma_d$ [$\epsilon/\sigma^2$] & $0.57 \pm 0.02$ & $0.82 \pm 0.03$ & $1.01 \pm 0.04$ \\
	\hline
	\end{tabular}
	\caption{\label{TabDroplet} Bulk properties of the polymer liquid at $k_B T=1.2\epsilon$ and $\rho_{\rm l}\approx 0.79\sigma^{-3}$.}
\end{table}

\subsection{Properties of the polymer liquid of the droplet}
\label{subsec:droplet}

Liquid droplets consist of $M=450, \cdots, 15328$ polymer chains of polymerization degree of $N=10,100$ and $200$ beads. Only the two last degrees of polymerization may involve entanglement effects \cite{Everaers04}, as the entanglement length is $N_{\rm e}\approx 64$ for this model. The interaction parameter in Eq.~\eqref{LJpot} was set to $\eps{dd}=1.0\epsilon$ for the polymer liquid of the drop. For the temperature $k_BT=1.2\epsilon$, the coexistence density of the fluid with its vapor is  $\rholo{l}\sim 0.79\sigma^{-3}$, while the vapor density can be negligibly small \cite{Pastorino07,Servantie08}.

Bulk properties of droplets are summarized in Tab.~\ref{TabDroplet}. The Rouse relaxation time $\tau_p$ is related to the self-diffusion coefficient $D=R_{\rm e}^{2}/\left(3\pi^2\tau_p\right)$, where $R_{\rm e}=\langle R_{\rm e}^{2}\rangle^{1/2}$ is the average end-to-end polymer distance. For the smallest chain length, $N=10$, this corresponds to very small values of the invariant degree of polymerization, $\bar{\mathcal{N}}\equiv(\rho_{\rm l} R^3_{\rm e}/N)^2\approx 14$.

The shear viscosity, $\eta_d$, has been computed using the reverse non-equilibrium molecular dynamics method \cite{MullerPlathe99, Tenney10}, which provides an efficient means compared to other (non)-equilibrium molecular dynamics methods \cite{Leonforte11}. To this end, one divides the system into slabs and exchanges momenta between the two slabs that are separated by half the system size. This creates a momentum flux between slabs and results in a linear velocity profile $v_{\parallel}(x)$. By varying the number of momentum swaps $N_{\rm s}= 3,15,60,120,300$, or $1200$ per integration step, one tunes the effective shear rate imparted onto the system. Simulation runs, which lasted between $5000\tau$ and $24000\tau$, were performed in a simulation cell containing $19160$ beads. Data were collected at half these times, when a steady state has been reached. The values obtained using this method are then extrapolated to vanishing momentum flux. Results are given in Tab.~\ref{TabDroplet} and are found to be in agreement with ones, for example, from equilibrium molecular dynamics simulations~\cite{Sen05,Leonforte11}.

Additionally, we computed the surface tension $\surft{d}$, using a slab geometry \cite{Nijmeijer88,Orea03}. The anisotropy of the pressure tensor yields the estimate $2A\surft{d}=V\left\{\langle P_{zz}\rangle - 0.5\left(\langle P_{xx}\rangle + \langle P_{yy}\rangle\right)\right\}$, where the cross section area is $A=L_x\times L_y$, and the factor of $2$ arises from the number of liquid/vapor interfaces in the simulation box with periodic boundary conditions.

\begin{figure}
\rsfig{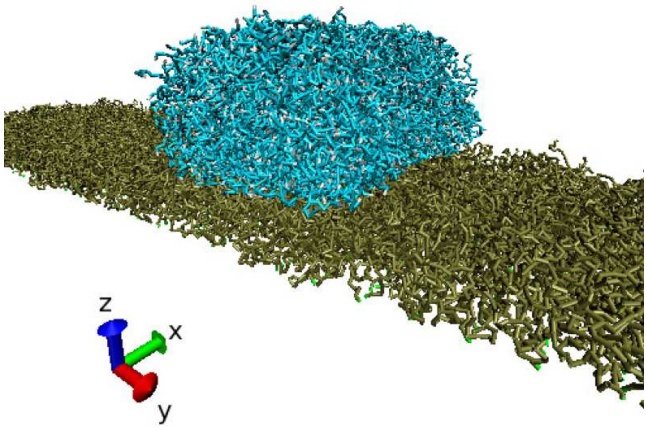}
\caption{(Color online) Sketch of the simulation set-up at $k_BT=1.2\epsilon$. Brush segments are colored in dark green, grafted ends are depicted in light green, while liquid droplet is colored in blue. Periodic boundary conditions are used in both $x$ and $y$ direction. The figure corresponds to a snapshot of the simulation box for a brush of grafting density $\rholo{g}R^2_{\rm e}=9.5$, and chain length $N=40$, while the droplet contains $M=1916$ chains of length $N=10$.}
\label{fig1}
\end{figure}

Except for Sec.~\ref{subsec:spherical}, cylindrical droplets are considered with a symmetry axis in the $x$ direction, along which periodic boundary conditions are applied. A snapshot of such a system is shown in Fig.~\ref{fig1}. In the following, brushes with an aspect ratio $A_r\equiv L_y/L_x=5$ will be used, also with periodic boundary conditions in $y$ direction, while impenetrable, flat surfaces limit the simulation box in $z$ direction. For large drops and small values of $L_x$, the cylindrical state is then an absolute minimum of the surface free energy, i.e. such a geometry stabilizes the cylindrical droplet against cap-shaped droplets, which will become energetically favorable for small droplets or larger system sizes, $L_{x}$. Cylindrical droplets are characterized by straight contact lines. Thus, we avoid strong finite-size effects which arise for spherical drops due to dependence of the length of the contact line on the droplet size and the concomitant contribution of the line tension \cite{Gretz1966,MacDowell02,Servantie08,Leonforte11,Berg10}.

\subsection{Characteristics of the soft, polymer brush}
\label{subsec:brush}

The deformable surface is modeled by an end-grafted brush of polymers comprised of $N=40$ beads. The interaction parameter in Eq.~\eqref{LJpot} is identical to the one used for the liquid droplet, namely $\eps{bb}=1.0\epsilon$. The grafting substrate is hard and impenetrable and the van-der-Waals interactions between the solid and the polymer brush and liquid are catered for by an integrated Lennard-Jones potential:

\begin{equation}\label{LJwall}
	{U^{\rm wall}_{\rm LJ}(r)} = \frac{2\pi\epsilon_{\rm wall}}{3}\Bigl[\Bigl(\frac{\sigma}{r}\Bigr)^9-\Bigl(\frac{\sigma}{r}\bigr)^3\Bigr] + \mbox{const} \qquad \mbox{for}\; z < 2.5\sigma\;,
\end{equation}

\noindent $U^{\rm wall}_{\rm LJ}=0$ for $z > 2.5\sigma$, and the constant in Eq.~\eqref{LJwall} is chosen such that $U^{\rm wall}_{\rm LJ}$ is continuous. We set $\epsilon_{\rm wall}=0.01\epsilon$ and $\sigma=1.0$. Since $\epsilon_{\rm wall} \ll k_{B}T$, the wall substrate is effectively repulsive. 

The first anchor segment of each polymer of the brush is irreversibly grafted and immobile. The anchor beads are regularly placed on a square grid in the $xy$-plane at a distance $z_0=0.8 \sigma$ above the substrate. The bond, ${\bf r}_{0}-{\bf r}_{1}$, between the immobile, anchor and the second, mobile bead of the polymer brush is represented by an anharmonic spring of the form $U_{\rm a}=\epsilon_{\rm a}\left({\bf r}_{1}-{\bf r}_0\right)^2.\left[\lambda^2-\left({\bf r}_{1}-{\bf r}_0\right)^2\right]^{-1}$~\cite{Rector94} with $\epsilon_{\rm a}=1.5\epsilon$ and $\lambda=0.8\sigma$. This anharmonic potential is softer than the FENE potential that is used for the other bonds and thus copes better with the strong forces on the bonds at the grafting substrate.
The grafting density, $\rholo{g}$ is defined by the number of polymer chains per unit area. We measure it in units of the end-to-end distance $R_{\rm e}=7.74\sigma$ of melt chains of length $N=40$. A value of $\rhored{g}\equiv\rholo{g}R^2_{\rm e}\sim\mathcal{O}(1)$ marks the crossover from mushrooms to a proper polymer brush. Systems with reduced densities of $6.5 \leq \rhored{g}\leq 57.4$ were investigated. This wide range of grafting densities represents different physical regimes for the deformable surface: A low grafting densities the surface is very soft; at high densities, the brush polymers are strongly stretched and the surface is significantly less deformable. 

First, we study the height, $H$, of the brush as a function of the grafting density $\rholo{g}$. The average height, $H$, is then defined as the first moment of the density distribution $H=\int z\Phi(z)dz/\int \Phi(z)dz$. We confirm that this quantity scales like $H\sim v N\rholo{g}^x$, where $v$ is the effective volume of a coarse-grained segment. The exponent, $x$, depends on the solvent quality. Since the temperature $k_{B}T=1.2 \epsilon$ corresponds to a polymer brush in a poor solvent, the brush is nearly incompressible and $v \sim 1/\rholo{l}$. Thus, the average height of the brush scales linearly with grafting density, $x=1$, in agreement with previous simulations~\cite{Grest99}.

Under bad solvent conditions, the top surface of a brush resembles the interface between a dense polymer liquid and its coexisting vapor of vanishingly small density. Conceptually, one has to distinguish between the "intrinsic" width of the brush surface and long-wavelength fluctuations of the local height of the brush, which are the analog of capillary waves on a free liquid-vapor interface. The long-wavelength fluctuations of the local height of the brush can be described in terms of an effective interface Hamiltonian \cite{Buff1965}:

\begin{multline}\label{effHamil}
	\mathcal{H}_{\rm cap}[z_{\rm int}(x,y)] = \frac{1}{2}\int_{A} {\rm d}^2{\bf r}_{\|}\Bigl(\surft{b}\left[\nabla z_{\rm int}\right]^2+\\
	\kappa_{\rm b}\left(z_{\rm int}-\bar{z}_{\rm int}\right)^2\Bigr)\;,
\end{multline}

\noindent where ${\bf r}_{\|}=(x,y)$ denotes the two lateral coordinates parallel to the grafting surface of area, $A=L_{x}L_{y}$. The variable $z_{\rm int}(x,y)$ is the instantaneous local position of the brush-vapor interface, and $\bar{z}_{\rm int}$ denotes the average height $H$. 
$\gamma_{b}$ denotes the surface tension of the brush and $\kappa_{b}$ characterizes the coupling between the average brush height and the grafting substrate. Since the brush is a slightly compressible liquid, fluctuations of the average brush height occur and their free-energy cost is proportional the the area of the grafting substrate and scale quadratically with the deviation from $H$. It is useful to consider the Fourier components of the local interface position,

\begin{multline}\label{interffluct}
	z_{\rm int}({\bf r}_{\|}) = \frac{a_0}{2} + \sum_k \left[c_c(\mathbf{q}_k)\cos{({\bf q}_k \cdot {\bf r}_{\|})}\right.+\\
	\left.c_s(\mathbf{q}_k)\sin{({\bf q}_k \cdot {\bf r}_{\|})}\right]\;,
\end{multline}

\noindent with wave vectors $\mathbf{q}_k=2\pi\left(k_x/L_x,k_y/L_y\right)$ and $k_x,k_y \in \mathbb{N}$. With this coarse-grained description in Eq.~\eqref{effHamil}, the equipartition theorem connects the spectrum of the brush-vapor interface fluctuations to the tension and deformability of the surface:

\begin{equation}\label{spectrum}
	\langle c^2_c(\mathbf{q})\rangle=\langle c^2_s(\mathbf{q})\rangle=    \frac{2k_BT}{A\left(\surft{b}\mathbf{q}^2+\kappa_{\rm b}\right)}\;.
\end{equation}

In order to evaluate Eq.~\eqref{spectrum}, the brush surface is divided into a grid of columns with lateral extension $\Delta_x=\Delta_y=1\sigma$. In each column, $(x,y)$ the brush-vapor interface position $z_{\rm int}(x,y)$ is defined as a Gibbs dividing interface \cite{Pastorino09,Muller1996} between the vapor and the brush melt. Data in a distance of $4\sigma$ away from the local interface position have been analyzed to minimize the effect of bulk-like density fluctuations of the compressible brush. About $10^3$ configurations with an aspect ratio $A_r=1$, separated by $5\tau$, were analyzed. The roughness $\delta\equiv\left[\langle\left(z_{\rm int}(x,y) - \bar{z}_{\rm int}\right)^2\rangle_{x,y,t}\right]^{1/2}$ of the brush, where the average is performed over all grid columns and time, was found to decrease (not shown) with an increase of $\rhored{g}$, as expected when the brush stretches. 

\begin{figure}
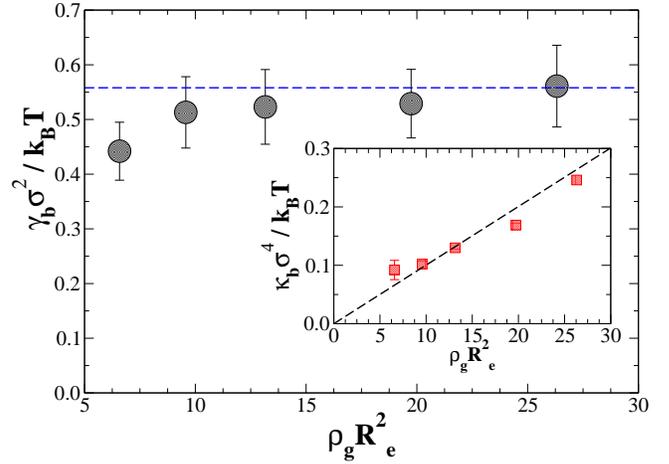

\rsfig{figure02.eps}
\caption{(Color online) Surface tension $\surft{b}$ of the $N=40$ beads per chain deformable surface as a function of the reduced grafting density $\rhored{g}=\rholo{g} R^2_{\rm e}$, extracted from the fluctuations spectrum, Eq.~\eqref{spectrum}. The horizontal dashed line is the liquid-vapor surface tension of a polymer melt with identical chain length. \emph{Inset:} Elastic constant $\kappa_{\rm b}$ of the brush surface as a function of the reduced grafting density. The data are compatible with a linear increase with the reduced grafting density as indicated by the dashed line.}
\label{fig2}
\end{figure}

The Fig.~\ref{fig2} shows the values of the normalized brush surface tension $\surft{b}\sigma^2/k_B T$ (main panel) and brush's elastic constant $\kappa_{\rm b}\sigma^4/k_B T$ (inset), extracted from the small $q$ behavior of the $y$-averaged Fourier components $c_c(q_x)=c_c(q_x,q_y=0)$, as a function of the reduced grafting density. For small grafting densities, the tensions $\surft{b}$ increases and saturates at larger $\rhored{g}$. Indeed, at intermediate grafting densities, the structure of the brush in a bad solvent is independent~\cite{Utz08} from $\rholo{g}$ and its surface resembles a liquid-vapor interface. Thus, we expect that the surface tension, $\surft{b}$, of the effective interface Hamiltonian is comparable to the surface tension of the free polymer liquid in coexistence with its vapor. For the parameters used in our simulation, $k_{B}T=1.2 \epsilon$ and $N=40$, the liquid-vapor interface tension is  $\surft{d}\sigma^2/\epsilon=0.67\pm0.02$ and this value is indicated by the horizontal line in Fig.~\ref{fig2}. Indeed, our simulation data are compatible with $\surft{b} \to \surft{d}$ in the limit of large grafting densities.

On the other hand, the elastic constant $\kappa_{\rm b}$, which is related to a uniform height variation, appears to dependent more strongly on the grafting density and the simulation results are compatible with a linear increase with the brush grafting density. If one assumed that each chain acts like an entropic spring with an elastic constant $k$, which is a reasonable assumption for a brush in contact with a melt of identical polymers \cite{Pastorino09}, that $\kappa_{\rm b}\simeq 6 k_B T\rhored{g}/R_{\rm e}^{4}$. For our system, we observe the same linear scaling with the grafting density but the pre-factor is about a factor $5$ larger. This difference presumably stems from the fact that our system is much less compressible than the brush-melt system studied in Ref.~\cite{Pastorino09}. In the brush-melt system the average height of the brush can fluctuate without a change of the segment density inside the brush -- free polymers of the melt in contact with the brush can enter or leave the brush to compensate for the variation of the density of brush segments. This mechanism is not available for the polymer brush in contact with vacuum, where a fluctuation of the average brush height entails a change of the polymer density and a concomitant free-energy penalty due to the equation-of-state of the polymer liquid.

\begin{figure}
\rsfig{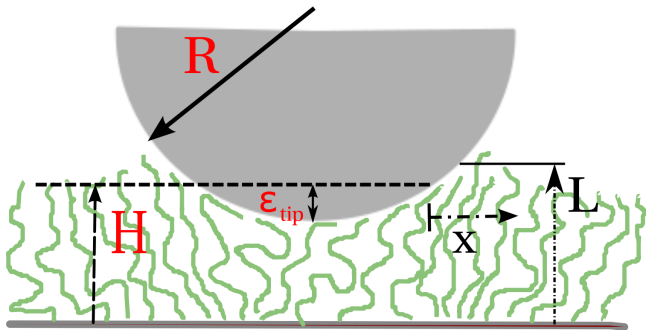}
\caption{(Color online) Sketch of the simulation analog of an AFM-indentation experiment and illustration of the quantities related to the measurement of the effect shear modulus, $G_{\rm b}$, of the brush. In the direction of the applied force, the brush responds like a liquid, while due to the constraints imposed by the grafting, the brush's response shows a resistance to the resulting lateral force.}
\label{fig3}
\end{figure}

Alternatively, the deformability of the surface can be characterized by its shear modulus, $G_{\rm b}$. Using computer simulations, this quantity is often difficult to access, while experimentally this quantity can be efficiently measured by indenting an AFM tip into the soft surface \cite{Halperin10,Akgun07}. Previous theoretical work by Fredrickson \emph{et al.} \cite{Fredrickson92} predicted that a molten polymer brush has an effective shear modulus, $G_{rm b}$, that causes the indentation behavior of a soft brush surface to deviate from that of a liquid. The shear modulus can be estimated by applying a lateral generic force, $F$, to the top of the brush surface, which ``shears'' the chains by displacing each chain end at the surface by a distance $x$, as depicted in Fig.~\ref{fig3}. Let $A_{\rm tip}$ be the lateral contact area of the indenter and $\epsilon_{\rm tip}$ the depth of the indentation. Following the qualitative considerations of Ref.~\cite{Fredrickson92},  one estimates that the free end of a typical brush molecule, which is located a distance $\sqrt{A_{\rm tip}}$ away from the center of the indenter, is laterally displaced by an amount $x \sim \sqrt{\rholo{g}A_{\rm tip}} \times \epsilon_{\rm tip}/(\sqrt{\rholo{g}}H)$, where the first factor accounts for the distance of the chain from the center of the indenter in units of the distance of the grafting points. The second factor is the lateral displacement of a chain at the center of the indented region. Such a lateral displacement gives rise to a lateral force, $f_{x} \sim k_{B}T x/R_{\rm e}^{2}$, on each chain and the total lateral stress generated by all chains underneath the indenter is $\sigma_{xz} \sim \rholo{g}f_{x} \sim \rholo{g} k_{B} T  x/R_{\rm e}^{2}$. Since the trajectories of the chains are approximately linear, the strain $u_{xz}$ is of the order $x/H$. Conceiving the brush as an incompressible, isotropic, elastic film \cite{Fredrickson92}, one can define an effective shear modulus of the brush via the relation $\sigma_{xz}=G_{\rm b}u_{xz}$, i.e. $G_{\rm b} \sim k_{B}T \rholo{g} H/R_{\rm e}$. Using $H \sim vN \sigma_{\rm g}$ and including the numerical pre-factor of the more detailed calculation in the framework of the Alexander-brush model \cite{Fredrickson92}, one arrives at:

\begin{equation}\label{FredGb}
G_{\rm b}  = 3 \frac{k_{B}T}{R_{\rm e}^{3}} \frac{vN}{R_{\rm e}^{3}} \rho^{\star2}_{\rm g}\;.
\end{equation}

Only recently, Fujii \emph{et al.} \cite{Fujii10} have shown, by measuring the power spectral density of monodisperse polystyrene brushes submitted to a tapping-mode AFM tip, that such theoretical predictions can describe the experimental data.

As illustrated in Fig.~\ref{fig3}, the AFM-like method is numerically mimicked by forcing a spherical indenter of radius, $R$, to compress the deformable surface. The lateral coordinates, $(x_{\rm tip},y_{\rm tip})$, of the indenter are fixed. The spherical indenter exerts a radial force of magnitude $F(r) = - K (r - R)^2$ on each segment, where the force constant is $K=100$ and $r$ stands for the distance of the segment to the center of the indenter. The force is repulsive for $r<R$, and $F(r) = 0$ for $r > R$. In the following, we use large tips, whose radii are greater than the brush height. This set-up corresponds to a compressive mode of indentation. Two types of tips are considered: \emph{(i)} a tip with a fixed radius $R\equiv R_{\rm fixed}$ and a center of mass $\mathbf{r}^{\perp}_{\rm tip}(t)$ that weakly varies in time as to describe the approach of the tip towards the brush surface, and \emph{(ii)} a tip with a slowly varying radius $R\equiv R_{\rm grow}(t)$ but a fixed center of mass $\mathbf{r}_{\rm tip}$. Such a weak time dependence is necessary if one wants to approach the fluctuating brush surface without perturbing it too much during the initial stage of penetration . In both cases, initial parameters ($R$, $\mathbf{r}^{\perp}_{\rm tip}(0)$ and $v^{\perp}_{\rm tip}$ for case  \emph{(i)}, and $\mathbf{r}_{\rm tip}$, $R_{\rm grow}(0)$ and $v^{\perp}_{\rm grow}$ for case \emph{(ii)}) are also adjusted in order to produce the same quasi-static protocol with respect to the penetration depth $\epsilon_{\rm tip}(t)\cong\epsilon_{\rm tip}$ (cf.~Fig.~\ref{fig3}).

The force $F_{\rm tip}$ required to push the spherical indenter into the brush at a given distance $\epsilon_{\rm tip}$ can be found in the limit $\epsilon_{\rm tip}\ll H\ll R$, these quantities being defined in Fig.~\ref{fig3}. Depending on the ratio between $H^2/R$ and the penetration depth $\epsilon$, one can distinguish two regimes: \emph{(i)} the strong penetration regime $(s)$, which is discussed in Ref.~\cite{Fredrickson92}, and \emph{(ii)} the weak penetration $(w)$ regime, which has also been considered in Ref.~\cite{Williams93}. The latter regime is equivalent to the situation where an infinitely hard sphere is pushed against a film of modulus $G_{\rm b}$. We only consider the latter regime and obtains the force $F_{\rm tip}$:

\begin{equation}\label{fw}
	F^w_{\rm tip} = \frac{16}{3} G_{\rm b} \epsilon^{3/2}_{\rm tip} R^{1/2} \,,\qquad \mbox{for} \quad \epsilon_{\rm tip} < \frac{H^2}{R}\;.
\end{equation}

Numerically, for the two types of tips, the average force exerted on the indenter is monitored and plotted for different penetration depths $\epsilon_{\rm tip}$. The dependence on $ \epsilon_{\rm tip}^{3/2}$ of Eq.~\eqref{fw} is then only observed for the lower grafting densities because the lateral displacements of the grafted chains are less constrained by their stretching. When $\rholo{g}$ increases, the compressibility of the surface decreases as the density inside the brush increases. The short range lateral perturbation of the brush in the vicinity of the inserted tip may thus be supplanted by additional long range deformation modes, which are not taken into account in the previous theory.

\begin{figure}
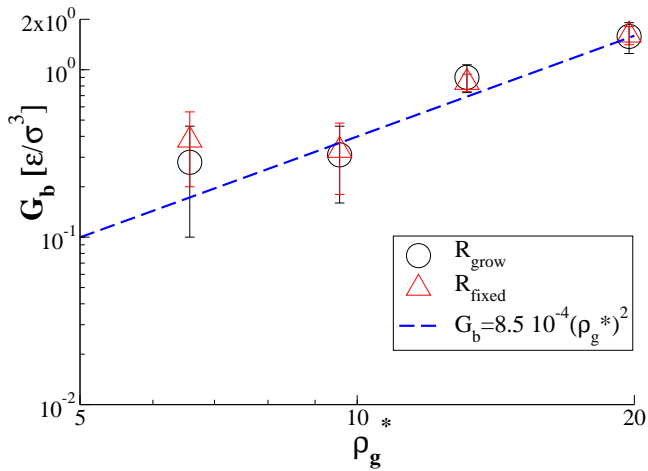

\rsfig{figure04.eps}
\caption{(Color online) Effective shear modulus of the brush as a function of the reduced grafting density $\rho^{\star}_{\rm g}$. Results are plotted for the two types of tips discussed in the text. The theoretical prediction $G_{\rm b} = 8.5 \cdot 10^{-4}\rho_{\rm g}^{\star 2} \epsilon/\sigma^{3}$, from Eq.~\eqref{FredGb}, is indicated by a dashed line.}
\label{fig4}
\end{figure}

In Fig.~\ref{fig4}, the effective shear moduli $G_{\rm b}$ of brushes with varying grafting densities is shown for the two types of tips. The data are compatible with a quadratic dependence of $G_{\rm b}$ on the grafting density, $\rhored{g}$, predicted by Eq.~\eqref{FredGb}. Using $v=1/\rholo{l}=1.27 \sigma^3$ and Eq.~\eqref{FredGb}, one obtains $G_{\rm b} = 8.5 \cdot 10^{-4}\rho_{\rm g}^{\star 2} \epsilon/\sigma^{3}$. The prediction qualitatively agrees with the simulation data. Possible reasons for the quantitative deviations are the rather short chain lengths and the finite compressibility of the polymer liquid in the simulations, as well as the strong stretching assumption invoked in the analytical calculation.

\section{Results}
\label{sec:results}

When a liquid is in contact with a solid surface, the balance of surface and interface tensions dictates the equilibrium properties of the liquid. Balancing the tensions parallel to the surface at the three-phase contact between the solid, the liquid and its vapor, one obtains Young's equation~\cite{Young1805} $\surft{LV}\cos{\theta_Y} + \surft{LS} = \surft{VS}$, where $\theta_Y$ is the equilibrium contact angle, and $\surft{LV},\,\surft{LS},\,\surft{VS}$ are the liquid-vapor, liquid-surface and vapor-surface surface tensions, respectively. If the surface is soft, then the forces that act at the three-phase contact line will also deform the surface and lift up the three-phase contact line \cite{Shanahan94,Carre95,Carre96}.

\begin{figure}
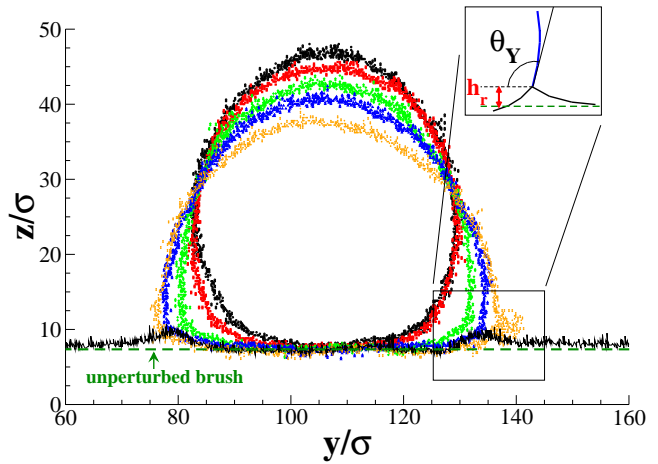

\rsfig{figure05.eps}
\caption{(Color online) Density contour plot of a polymer drop ($N=10$) wetting a brush ($N=40$) for various compatibility parameters $\eps{bd}=0.3\epsilon$ (black) to $\eps{bd}=0.8\epsilon$ (orange) between the brush and the droplet. The reduced grafting density of the polymer brush is $\rhored{g}=13.1$ and the droplet contains $M=3832$ polymers. The dashed line represents the average free surface position of the unperturbed brush. The lifting-up of the contact line and the formation of a ridge is visible. \emph{Inset:} Sketch of the procedure used to extract the static wetting contact angle, $\theta_Y$, and the strength  of the lifting-up, $h_{r}$, here for a compatibility $\eps{bd}=0.6\epsilon$.}
\label{fig5}
\end{figure}

In our simulations, making the molecules of the brush and the polymer liquid slightly incompatible, we can independently control the contact angle and the deformability of the polymer brush. The brush is the softer the higher the grafting density $\rholo{g}$ is. Therefore, quite low grafting densities are used in the following. The different wetting properties are explored as a function of $\eps{bd}$: a value $\eps{bd}/\epsilon=1$ leads to complete wetting, i.e. a thick polymer film spreads on the brush, while smaller values of $\eps{bd}$ give rise to polymer droplets with a finite contact angle. We will refer to $\eps{bd}$ as the compatibility parameter.

\subsection{Wetting of a cylindrical droplet}
\label{subsec:cylindrical}

Both, in experiments~\cite{Fondecave98,Brochard02} and in computer simulations~\cite{Fujii10,Dimitrov10,Saville77,Milchev01_1,Milchev01_2,Weijs11,DeConinck08}, the determination of the apparent contact angle remains a delicate task. Commonly, for flat surfaces, an apparent, local contact angle is graphically extracted at the contact line, from the value of the angle between an extrapolated tangent plane to the droplet's surface and the plane of the flat surface. Since we deal with deformable surfaces, the task is even more delicate, and we have opted for a carefully estimate from the density contour plots, such as ones depicted in Fig.~\ref{fig5}. The method of determining the contact angle is illustrated in the inset of this figure. 
In order to obtain an accurate estimate of the contact angle, the static properties of the brush/droplet systems were evaluated by collecting $10^3$ snapshots over simulation runs of $500\tau$. In Fig.~\ref{fig5}, the density contours are plotted for a liquid droplet of $M=3832$ polymers with $N=10$ beads per chain. The drop is in contact with a brush surface of grafting density $\rhored{g}=13.1$, and contours are plotted for various compatibilities $0.3\leq\eps{bd}/\epsilon\leq 0.8$.

As expected, it appears in Fig.~\ref{fig5} that the droplet deforms its shape when the compatibility with the surface increases, leading to \emph{(i)} a contact angle variation, and \emph{(ii)} to the appearance of a lifting-up of the surface at the three-phase contact line.

\begin{figure}
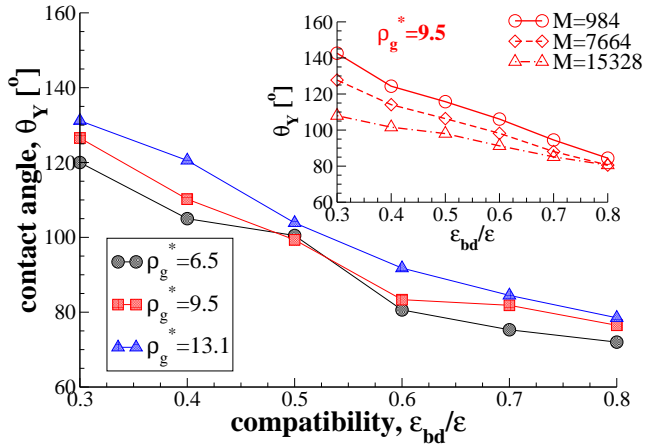

\rsfig{figure06.eps}
\caption{(Color online) Equilibrium contact angle, $\theta_Y$, between the brush and the droplet as a function of compatibility, $\eps{bd}/\epsilon$. Main panel: Contact angle for $M=1916$ polymers of length $N=10$ and the three lowest brush grafting densities $\rhored{g}=\rholo{g} R^2_{\rm e}$. The contact angle $\theta_Y$ slightly increases with $\rhored{g}$. \emph{Inset:} Contact angle for a fixed $\rhored{g}=9.5$ and increasing the number of polymer chains in the droplet. The contact angle, $\theta_Y$, slightly decreases with the increase of $\rhored{g}$ at fixed $\eps{bd}/\epsilon$. }
\label{fig6}
\end{figure}

In Fig.~\ref{fig6} the dependence of the contact angle on the compatibility $\eps{bd}$ is presented. Droplets with a  fixed number of polymers and a degree of polymerization $N=10$ wet brushes at intermediate grafting densities. As one increases the compatibility, the contact angle decreases~\cite{Dimitrov10}. It also appears a slight dependence on $\rholo{g}$, which becomes even smaller as one increases the grafting density (not shown), because at very high grafting densities the surface of the brush in a bad solvent is largely independent from the grafting density, $\rholo{g}$. Similarly, also brush-droplet interface is rather independent of $\rholo{g}$ for dense brushes. On the other hand, at low grafting densities, the decrease of the contact angle at fixed $\eps{bd}$ reflects the intricate changes of the brush-melt interface in response to changing $\rholo{g}$. We note that a value of $\eps{bd}/\epsilon\approx 0.6$ is sufficient to achieve a contact angle of $90\degree$, which is expected to result in the largest deformation of the brush at the three-phase contact line. 

The deformation of the surface and the building of a wetting ridge at the three-phase contact line are dictated by a competition between the minimization of the droplet's surface and the deformation of the brush surface and brush-melt interface, respectively. In the inset of Fig.~\ref{fig6}, the reduced grafting density of the brush is fixed to $\rhored{g}=9.5$, while the number, $M$, of polymer chains in the droplet increases. We observe that at fixed compatibility, $\eps{bd}/\epsilon$, the contact angle decreases when the size of the droplet increases. Hence, smaller drops seem to balance surface, adhesion and elastic deformation energies in a different manner than larger drops. 

\begin{figure}
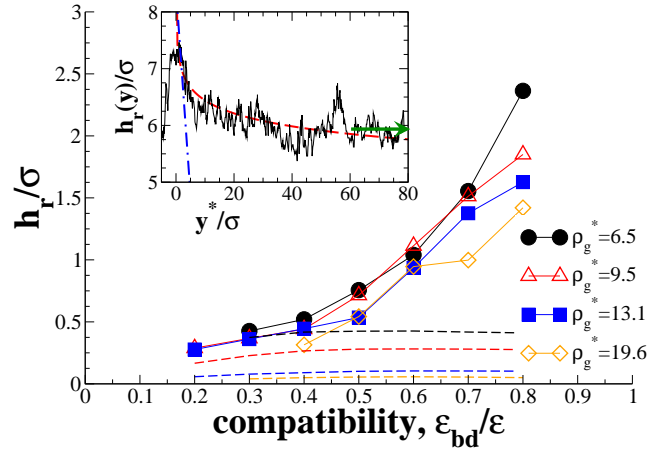

\rsfig{figure07.eps}
\caption{(Color online) Dependence of the height $h_{\rm r}$ of the ridge at the three-phase contact line as a function of the polymer drop ($N=10$) and brush ($N=40$) compatibility $\eps{bd}/\epsilon$. The drops contain $M=1916$ polymers, and low to intermediate brush grafting densities are considered. The dashed lines are fits using Eq.~\eqref{wetRidge} with contact angles from the Fig.~\ref{fig6}. \emph{Inset:} wetting ridge profile for a $M=3832$ polymer chains droplet with a chain length $N=10$. The droplet wets a brush of grafting density $\rhored{g}=9.5$ with a compatibility $\eps{bd}/\epsilon=0.7$. The rescaled distance from the origin $y^{\star}/\sigma$, is defined at the three-phase contact line. The Eqs.~\eqref{wetRidge}  and \eqref{wetApex} are plotted, respectively with dashed and mixed-dashed lines, for $\surft{d}\equiv\surft{LV}$, $\surft{BV}$ and $G_{\rm b}$ taken from the Tab.~\ref{TabDroplet}, Fig.~\ref{fig2} and Fig.~\ref{fig4}, also respectively. The arrow marks the average brush height far away from the ridge.}
\label{fig7}
\end{figure}

In Fig.~\ref{fig7}, the height $h_{\rm r}$ of the lifting up at the three-phase contact line is presented as a function of the compatibility $\eps{bd}/\epsilon$. Data are extracted from the density contour plots, such as the ones drawn in Fig.~\ref{fig5}. Results are only shown for the $N=10$ beads droplets that contain $M=1916$ polymer chains, the trend is the same for the other droplet sizes and degrees of polymerization, as also noted in Ref.~\cite{Carre95}. We consider small to intermediate brush grafting densities. Because the main deformation is localized at the edge of the brush-droplet interface, its strength is of the order of the monomer size $\sigma$, and it slightly increases when the grafting density decreases and the surface becomes more deformable. We observe a weak dependence on the brush grafting density $\rhored{g}$, in particular because the deformation is rather localized in the vicinity of the three-phase contact line and the narrow brush-melt/brush-vapor interface. We have also observed (not shown) that the vertical displacement $h_{\rm r}$ at the contact line has a magnitude that depends on the droplet size~\cite{Lester61,Das2011} as $\sim \left(\surft{d}/G_{\rm b}\right)\ln{\left(R_{\rm d}/\sigma\right)}$, where $\surft{d}$ and $R_{\rm d}$ are respectively the droplet surface tension and its radius. This dependence has been checked for an estimate of $R_{\rm d}$ from density contours, as well as from a rough estimate for cylindrical droplets, i.e. $R_{\rm d}\propto\sqrt{N\times M}$.

It appears that the height, $h_{\rm r}$, of the ridge increases when the compatibility increases to $0.8$. This behavior is surprising because the contact angle decreases with $\eps{bd}/\epsilon$, and the liquid-vapor interface tension acts more tangentially to the surface. In the limit $\eps{bd}/\epsilon\rightarrow 1$, the liquid-vapor interface makes zero (wetting) or a very shallow angle (autophobicity) with the brush surface. Starting from a droplet geometry with a large contact angle, thus, we expect a deformation of the contact line with an increase of the wetting ridge to a maximum value, followed by a decrease towards zero, as we increase the compatibility. Due to the softness of the surface that also slows down droplets spreading, we were not able to computationally access this time scale.

The resulting wetting ridge of the soft surface has a height $h_{\rm r}(y)$ that decreases from the three-phase contact line position. It has been theoretically proposed by Carre \emph{et al.}~\cite{Shanahan94,Carre95,Carre96} and verified in experiments, that $h_{\rm r}(y)$ has an asymptotic behavior of the form:

\begin{equation}\label{wetRidge}
 h_{\rm r}(y)\approx h_{\rm r}(0)\ln{\left(\frac{d}{y}\right)}\text{,   where $h_{\rm r}(0)=\frac{\surft{d}\sin{\theta_Y}}{2\pi G_{\rm b}}$}\;,
\end{equation}
 
 \noindent where $d$ represents the distance away from the three-phase contact at which no displacement occurs within the surface, $\surft{d}$ the droplet surface tension, and $G_{\rm b}$ the shear modulus of the deformable solid. The height at the origin, $h_{\rm r}(0)$, is taken as the reference position of the three-phase contact line. In the Fig.~\ref{fig7}, we plot the this height as a function of the compatibility $\eps{bd}/\epsilon$, using the associated contact angles from the Fig.~\ref{fig6}. First, the theoretical prediction is found to underestimate the strength of the lifting-up of the contact line measured in the simulations. This prediction also depicts a maximum at $\theta_Y=\pi/2$, before decreasing at smaller angles; a trend that is not observed in our simulations, where $h_{\rm r}(0)$ continuously grows with the decrease of $\theta_Y$. On the other hand, we find that the logarithmic dependence on the distance from the contact line is compatible with our simulations data, as shown in the inset of Fig.~\ref{fig7}, where the height of the ridge at the contact line is the one measured in our simulations.

Eq.~\eqref{wetRidge} predicts a logarithmic dependence that becomes singular at the three-phase contact line. It has been recently proposed~\cite{Jerison2011} to model the shape in the vicinity of the apex by a cusp:

\begin{equation}\label{wetApex}
h_{\rm r}(0)-h_{\rm r}(y)=\frac{\surft{LV}}{\surft{BV}}\sin{\theta}\mid y\mid\;,
\end{equation}

\noindent which then assumes that in the vicinity of the three-phase contact line, the surface elasticity does not affect the shape of the cusp. Such a behavior is plotted and compared to the ridge profile in the inset of the Fig.~\ref{fig7}, where a linear dependence is indeed compatible with the simulation  at small distances from the apex, $d \lesssim 5\sigma$.

\subsection{Wetting of a spherical droplet}
\label{subsec:spherical}

In this section, we study spherical, cap-shaped droplets in order to investigate the line tension on a soft, deformable substrate. 

\subsubsection{Line tension effect}
\label{subsubsec:sph_linetension}

When one deals with spherical droplets, one also has to account for the curvature $\kappa$ of the three-phase contact line, which may have a significant contribution on the wetting properties of small enough drops. This effect is commonly represented via the line tension, $\tau$, that acts over the circumference length $2\pi a$ of the wetting sphere-cap shaped droplet, and modifies Young's equation according to:

\begin{equation}\label{YoungSph}
	\cos{\theta_Y} = \cos{\theta_{\infty}} - \frac{\tau}{\surft{d}}\kappa\;,
\end{equation}

\noindent with $\surft{d}$ the droplet surface tension with its vapor, and $\theta_{\infty}$ the contact angle of a droplet with zero curvature, typically either a macroscopic droplet of a cylindrical one that has a straight contact line. The contact angles $\theta_Y$ are extracted from the equilibrium profiles of spherical droplets ranging from $M=984$ to $M=30656$ polymer chains of length $N=10$. The drops wet different deformable surfaces of same grafting densities $\rholo{g}$ than the  ones already used in the Sec.~\ref{subsec:cylindrical}. The brushes contain between $M=3969\,,(L_x=L_y=212\sigma)$ and $M=11881\,,(L_x=L_y=352\sigma)$ grafted chains of length $N=40$, hence the lateral dimensions are large enough to avoid finite size effects.

\begin{figure}
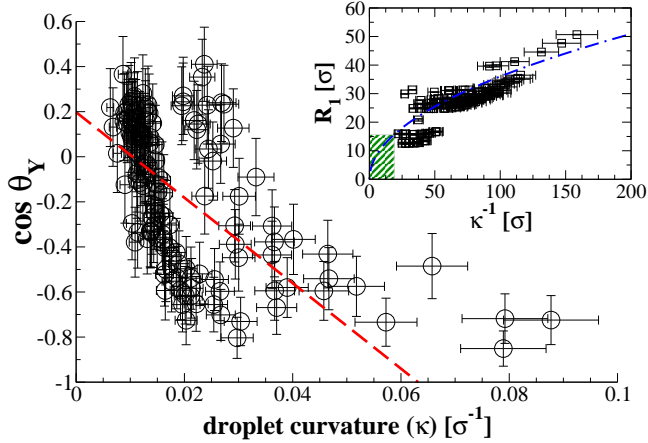

\rsfig{figure08.eps}
\caption{(Color online) Cosine of the spherical droplet contact angle, $\theta_Y$, as a function of curvature, $\kappa$, at the three-phase contact line. The slope of the dashed red line corresponds to the characteristic length scale, $\tau/\surft{d}\approx 19\sigma$. The same kind of curves can be observed experimentally~\cite{Berg10} for nanometer-size sessile droplets. \emph{Inset:} droplet radius $R_1$ \emph{vs} the contact radius $\kappa^{-1}$, measured independently from the simulations. The blue line is a square-root fit to the spherical-cap approximation, which is more accurate for large and slightly deformed droplets. This gives an estimate of the critical droplet radius, $R_c\approx 15.7\sigma$, below which line tension effects become important.}
\label{fig8}
\end{figure}

Results for $\cos{\theta_Y}$ versus the curvature~\cite{Berg10} of the three-phase contact line $\kappa$ are plotted in Fig.~\ref{fig8}. The dashed red line is the best fit that assumes a behavior according to the Eq.~\eqref{YoungSph}, with $\theta_{\infty}=78.61\degree\pm5.94\degree$ and the characteristic length scale $\tau/\surft{d}=18.96\pm 6.04\sigma$. In the limit of large spherical droplets for which the contact line term is not relevant, $\theta_{\infty}$ is then of same order than $\theta_Y$ measured for cylindrical droplets of equivalent surface contact radius (inset of Fig.~\ref{fig6} for the largest system).

Therefore, one can relate the contact radius $\kappa^{-1}$ to the apparent droplet radius $R_1$ measured independently. For large and slightly deformed droplets, the spherical cap approximation connects the contact radius $\kappa^{-1}$ to the square-root of the apparent droplet radius $R_1$. This is plotted in the inset of Fig.~\ref{fig8} in order to fit the measured data. Using the threshold value $\tau/\surft{d}$, one can estimate the critical droplet size, $R_c\approx 15.7\sigma$, below which the line tension may affect, at equilibrium, the wetting properties of spherical droplets~\cite{Milchev01_2}. In Fig.~\ref{fig8}, this corresponds to the shaded region. To reduce significant curvature effect at the contact line as well as dependence of the contact angle with the droplet size and brush/droplet compatibility, one then has to consider droplets with an higher radius.

One way to quantify the line tension effect consists in using a schematic model that incorporates the free-energy cost stemming from the curvature of the contact line. To this end, we use an effective Hamiltonian~\cite{Muller2011,Muller2002}, which also allows us to qualitatively study the interplay between the surface and interfacial tensions, the line tension, and the softness of the substrate that dictates the droplet shape and the wetting ridge.
The effective interface Hamiltonian, $\mathcal{H}$, accounts for the free energies of the droplet's surfaces, the elastic properties of the deformable surface, and the line tension. For a three dimensional system, $\mathcal{H}$ takes the form:

\begin{equation}\label{EffHamil}
\mathcal{H} = \sum_I \gamma_I A_I + \tau L + \lambda V_d + \beta H_b\;,
\end{equation}

\noindent where the sum runs over the different interfaces $I$ of area $A_I$,  $L$ is the length of the three-phase contact line, and $\lambda$ is a Lagrange multiplier for the fixed volume constraints of the droplet. The brush-like surface is modeled by vertical harmonic springs of stiffness $k$ proportional to the grafting density, as already discussed in the Sec.~\ref{subsec:brush}. The stiffness is adjusted in order to approximate the one of our soft surfaces, and using Eq.~\eqref{FredGb}, we get $k=15 G_{\rm b}/(R^2_{\rm e}\rholo{g})$. This leads to the additional energy contribution $H_b=\int k/2(\rho-\rho_0)^2 d{\bf r}$ in the total Hamiltonian \eqref{EffHamil}, which can be discretized in local densities $\rho_{i_{\rm cell}}$ of spring segment lengths. Therefore, one can write $H_b\equiv\sum_{i_{\rm cell}} k\Delta V_{\rm cell}/2\left(\rho_{i_{\rm cell}}-\rho_0\right)^2$, where $\rho_0$ is the initial spring segment lengths density in cells, $\Delta V_{\rm cell}=\left(L_{y_{\rm cell}}\times L_{z_{\rm cell}}\right)$ and the edge length $L_{\alpha_{\rm cell}}$ of the cell is half the spring equilibrium length. Other interfaces are also modeled by springs of stiffness inversely proportional to the spring equilibrium length. The Eq.~\eqref{EffHamil} is then minimized, using a Monte-Carlo scheme, for different initial droplet states $\theta_0$ and radii $R_p$, and for a spreading coefficient $S\equiv\surft{BV}-\surft{BL}-\surft{LV}$ that is independently tuned in the partial wetting regime, i.e. $S < 0$.

\begin{figure}
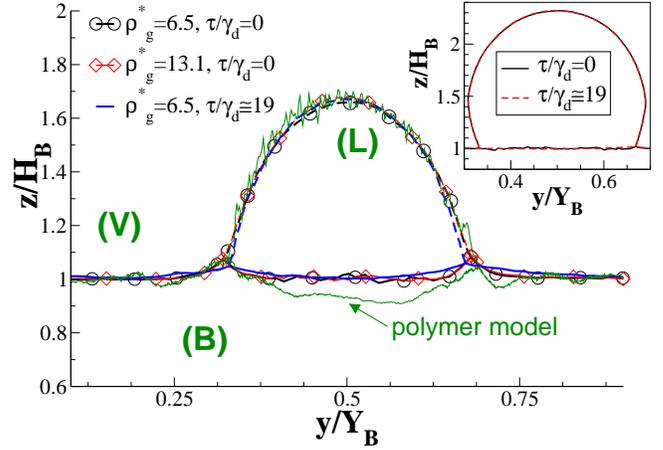

\rsfig{figure09.eps}
\caption{(Color online) $B/L/V$ Interfaces in a minimalistic droplet/soft-surface model. The shapes are obtained by minimizing the Eq.~\eqref{EffHamil}, in which surface tensions and brush elasticity are mapped onto simulation systems. An initial droplet of radius $R_p < R_c$ is considered, and the interface Hamiltonian is minimized including or not the line tension term $\tau/\surft{d}$ from the Eq.~\eqref{YoungSph}. Results from the polymer model are also superimposed for the same size regime, contact angle and a brush grafting density $\rhored{g}=13.1$. \emph{Inset:} a larger droplet of initial radius $R_p > R_c$ is considered, for a contact angle $\theta_0=2\pi/3$ and an effective grafting density $\rhored{g}=6.5$.}
\label{fig9}
\end{figure}

In the main panel of Fig.~\ref{fig9}, results from minimizations of Eq.~\eqref{EffHamil} are plotted, in rescaled units, for an initial droplet of radius less than $R_c$ and $\theta_0=\pi/2$, and for a value of $\surft{BL}$ that fulfills the Young's equation for $\theta_0$. For quantitative comparison, rescaled data from MD simulation on the polymer model are superimposed, and for the same size regime, contact angle and surface softness of $\rhored{g}=13.1$ . For the minimization procedure, we have also considered two surfaces that mimic the ones we use in our simulations by adjusting the brush stiffness $k$ accordingly. As in the Fig.~\ref{fig7}, the strength of the ridge slightly depends on the surface compliance, at least in the range of stiffnesses we considered. Its strength and trend also roughly follow the ones given in the Fig.~\ref{fig7} for the current value of the contact angle, as well as compared to MD data on the polymer model. We also observe a decrease of the ridge when a line tension of same order than the one we measured in our simulations is included in the Eq.~\eqref{EffHamil}, which therefore increases the equilibrium contact angle value, as also observed in Ref.~\cite{Milchev01_2}. In the inset of the Fig.~\ref{fig9}, the initial droplet radius $R_p$ is larger than $R_c$, while the initial angle $\theta_0$ is equal to $2\pi/3$, the surface softness mimics a grafting density $\rhored{g}=6.5$, and Eq.~\eqref{EffHamil} is minimized with and without the line tension term. In this droplets size regime, the line tension strength is supposed to not affect the equilibrium contact angle, which is indeed observed.

\subsubsection{Wetting regimes}
\label{subsubsec:sph_wettregime}

\begin{figure}
\rsfig{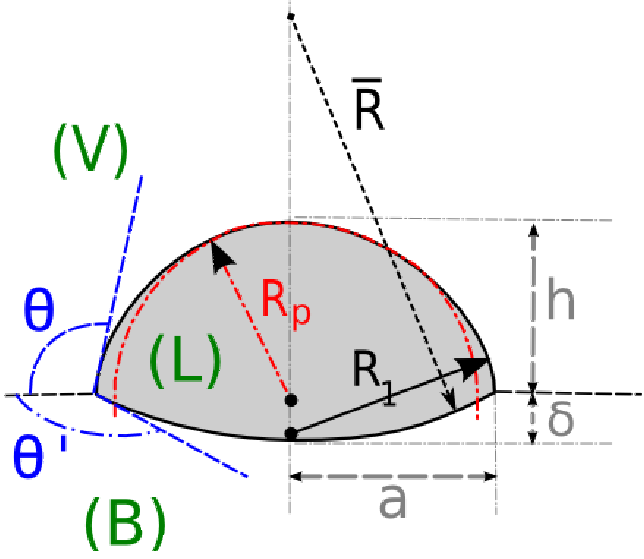}
\caption{(Color online) Two dimensional schematics of the spherical droplet model (L) at the interface between a surface (B) and the vapor (V). Also drawn are the initial droplet radius $R_p$, its upper ($R_1$) and lower ($\overline{R}$) spherical cap approximation radii, the projected contact radius $a$, the penetration depth $\delta$ and the visible droplet height $h$, as well as the two relevant contact angles $\theta$ and $\theta'$. }
\label{fig10}
\end{figure}

In the following, we aim to evaluate how the characteristics of the droplet and the soft surface, can dictate the equilibrium shape of the droplet. We first estimate the free energy that can be gained, by a droplet, by deforming the surface. We consider two limiting case, \emph{(i)} wetting of a droplet over a hard surface of shear modulus $G_{\rm b}\rightarrow\infty$, and \emph{(ii)} wetting over a liquid surface ($G_{\rm b}\rightarrow 0$) and for which the surface deformation $\delta$ is small compared to the droplet height $h$ (see Fig.~\ref{fig10} for a sketch of the model). 

In the case \emph{(i)} and using notations from the Fig.~\ref{fig10} with $\delta\rightarrow 0$, the balance of tangential forces at the three-phase contact line gives:

\begin{equation}\label{DeltaEi1}
\Delta E_{(i)}=\pi a^2\left[\left(\surft{BL}-\surft{BV}\right)+\surft{LV}\left(1 - \frac{h}{R_1}\right)\right]\;,
\end{equation}

\noindent where the $\surft{ij}$ are the surface tensions of the different interfaces drawn in the Fig.~\ref{fig10}. Using the spherical cap approximation, the spherical cap radius writes $R_1=\left(a^2+h^2\right)/2h$, and one obtains for the free energy cost:

\begin{equation}\label{DeltaEi2}
\Delta E_{(i)}\simeq \pi a^2\left[\surft{BL}-\surft{BV}+\surft{LV}\frac{a^2-h^2}{a^2+h^2}\right]\;.
\end{equation}

In the case \emph{(ii)}, the lense-shaped droplet of the Fig.~\ref{fig10} involves a free energy cost, when it wets a liquid, of the form $\Delta E_{(ii)}=-\pi a^2S+\pi\Bigl[\surft{LV}h^2+\surft{BL}\delta^2\Bigr]$, where the spreading parameter writes $S\equiv\surft{BV}-\surft{BL}-\surft{LV}=\surft{LV}(\cos\theta_Y-1)$. Therefore, in the spherical cap approximation, the difference $\Delta E=\Delta E_{(ii)}-\Delta E_{(i)}$ provides an estimate how much free energy can be gained by deforming the surface, and one obtains:

\begin{equation}\label{DeltaE}
	\Delta E=\pi\left\lbrace\surft{BL}\delta^2+\surft{LV}h^2\frac{\left(3a^2+h^2\right)}{a^2+h^2}\right\rbrace\;.
\end{equation}

Then, we estimate the free energy cost of deforming a soft surface that is mimicked by a polymer brush, thus with the constraint of attachments on one chain ends. Using notations from the Fig.~\ref{fig10}, this one involves the formation part of the free energy of the drop, and the one of a round AFM tip of radius $\overline{R}$ that gives rise to the Eq.~\eqref{fw}, namely $U_{el}=(32/15)G_{\rm b}\overline{R}^{1/2}\delta^{5/2}$. For the formation of the drop, the free energy is split in the energy cost of displacing an area $\pi a^2$ of $B/V$ interface, and the one that comes from the upper ($R_1$) and lower cap radii ($\overline{R}$) and their related interfaces. The final free energy is then written in the spherical cap approximation, which gives:

\begin{multline}\label{DeltaF}
	\Delta F=\frac{32}{15\sqrt{2}}G_{\rm b}\left(a^2+\delta^2\right)^{1/2}\delta^2 - \pi a^2 S +\\ \pi\left(\surft{LV}h^2+\surft{BL}\delta^2\right)\;,
\end{multline}

\noindent where $G_{\rm b}$ is the shear modulus of the soft surface. Thus, Eq.~\eqref{DeltaF} approximates the rubber surface as a liquid-like surface that has an elastic response under deformation. It combines both the droplet deformation under wetting and its effect on deforming the surface. 

By comparing Eqs.~\eqref{DeltaE} and~\eqref{DeltaF}, we note that the asymptotic limit $\Delta F/ \Delta E\gg 1$ corresponds to a hard surface for which the free energy cost is prohibitive, and for which the Young's equation dictates the droplet shape. The opposite limit $\Delta F/\Delta E\rightarrow 0$ corresponds to a very cheap cost of deformation, and the lens-shaped droplet wets a liquid. At intermediate values, $\Delta F/\Delta E < \pm 1$, a \emph{concurrent coupling} appears between the elastic properties of the drops and the surface, and therefore, a more intricate behavior dictates the droplet shape. One also should note that for $\Delta E>0$ it is more favorable for the drop to wet a solid, while for $\Delta E <0$, it costs more.

In order to extract a cross-over length, and following Refs.~\cite{Fredrickson92,Long96} for a small perturbation of the brush surface, we approximate the elastic free energy $U_{el}\approxeq A_b G_{\rm b}(\delta^2 a^4)/\langle H\rangle^3\equiv\widetilde{A}_b G_{\rm b} a^6/\langle H\rangle^3$, where $\langle H\rangle$ is the average brush height and $\widetilde{A}_b$ is given in the Eq.~\eqref{third}. Using notations of the Fig.~\ref{fig10}, we also rewrite Eq.~\eqref{DeltaF} in a more compact and completely equivalent fashion:

\begin{subequations}\label{DeltaFNew}
\begin{align}
\Delta F &=\widetilde{A}_b G_{\rm b} a^6/\langle H\rangle^3+A_d\surft{BV}\pi a^2\label{second}\\
A_d &=\surft{BV}^{-1}\left[-S+\surft{LV}\tan^2{\left(\theta/2\right)}+\surft{BL}\tan^2{\left(\theta'/2\right)}\right]\label{third}\\
\text{and, } \widetilde{A}_b &=(3\pi/2)\tan^2{\left(\theta'/2\right)}\;.\label{fourth}
\end{align}
\end{subequations}

As we are seeking for a preferable droplet scale for which wetting over a deformable surface is stable and favoring, we then minimize the free energy $\Delta F$ with the constraint that the droplet volume is constant. For this purpose, one defines the Lagrange function:

\begin{subequations}\label{MinDeltaF}
\begin{align}
	L\left(a,\theta,\theta',\lambda\right)&=\Delta F\left(a,\theta,\theta'\right)-\lambda\left[V_d\left(\theta,\theta'\right)-V_p\right]\label{MDFa}\\
	\text{and, } V_d\left(\theta,\theta'\right)&=A_v a^3\label{MDFb}\\
	&=(\pi/6)\lbrace 3\tan{\left(\theta/2\right)}+\tan^3{\left(\theta/2\right)}\nonumber\\
	&\quad+3\tan{\left(\theta'/2\right)}+\tan^3{\left(\theta'/2\right)}\rbrace a^3\nonumber\;.
\end{align}
\end{subequations}

\noindent where, in Eq.~\eqref{MDFa}, $\lambda$ is a Lagrange multiplier and $V_p\equiv\left(4\pi R_p^3\right)/3$ is the initial droplet volume, using notations from the Fig.~\ref{fig10}. The principle of least energy and the condition of $\partial L/\partial \theta=0$ yield to

\begin{equation}\label{MinVconstraint}
	\lambda=4a^{-1}\surft{LV} \tan{\left(\theta/2\right)} \cos^2{\left(\theta/2\right)}\;.
\end{equation}

\noindent The condition $\partial L/\partial a=0$, in which the result from Eq.~\eqref{MinVconstraint} is inserted, gives:

\begin{equation}\label{EqualMin1}
	\frac{\widetilde{A}_b G_{\rm b} a^4}{2 A_v \surft{LV} \langle H\rangle^3} + \frac{\pi A_d \surft{BV}}{6 \surft{LV} A_v}=\tan{\left(\theta/2\right)} \cos^2{\left(\theta/2\right)}\;.
\end{equation}

\noindent Finally substituting Eq.~\eqref{MinVconstraint} in the condition $\partial L/\partial \theta'=0$, one obtains:

\begin{multline}\label{EqualMin2}
	\tan{\left(\theta'/2\right)} \cos^2{\left(\theta'/2\right)}\left[\frac{3 G_{\rm b} a^4}{2\surft{LV}\langle H^3\rangle} + \frac{\surft{BL}}{\surft{LV}}\right]=\tan{\left(\theta/2\right)} \\
	\times\cos^2{\left(\theta/2\right)}\;.
\end{multline}

\noindent In view of Eq.~\eqref{EqualMin1}, one can note that the right-hand side is of $\mathcal{O}(1)$, which can be fulfilled on the left-hand side by requiring the condition:

\begin{equation}\label{EqualMin3}
\frac{\widetilde{A}_b G_{\rm b} \left(a^{\star}\right)^4}{2\langle H^3\rangle}\sim\frac{\pi A_d\surft{BV}}{6}\;,
\end{equation}

\noindent where $a^{\star}$ appears to be a characteristic scale. Using the relation $\langle H\rangle\sim \upsilon N\rholo{g}=\upsilon\rhored{g}N/R_e^2$ discussed in Sec.~\ref{subsec:brush}, one finally arrives to the expression of a cross-over length scale at which the free energy cost of deforming a soft surface with respect to the droplet shape is minimal: 

\begin{equation}\label{xi}
a^{\star}\equiv\xi=\left(\frac{N}{R^2_{\rm e}}\right)^{3/4}\left[\frac{\pi A_d\surft{BV}\left(\rhored{g}\right)^3}{3\widetilde{A}_b G_{\rm b}}\right]^{1/4}\;,
\end{equation}

\noindent where $N$ and $R^2_e$ are respectively the number of monomers in a brush chain, and the end-to-end average radius for this length $N$. This scale marks the border between a domain where droplets of radius lower than $\xi$ have an equilibrium shape dictated by a surface tension/elasticity coupling, to another domain for larger droplets where elasticity dominates. 

To evaluate Eqs.~\eqref{DeltaE}, \eqref{DeltaF}, and \eqref{xi} over different regimes,  the size of the droplets was controlled by increasing the number $M$ of polymer chains inside the droplet, i.e. $R_p\propto M^{1/3}$, while different wetting regime were considered by either tuning the brush/droplet compatibility $\eps{bd}/\epsilon$ and the brush grafting density $\rhored{g}$. In any case, only droplets of radius $R_p>R_c$ were simulated.

\begin{figure}
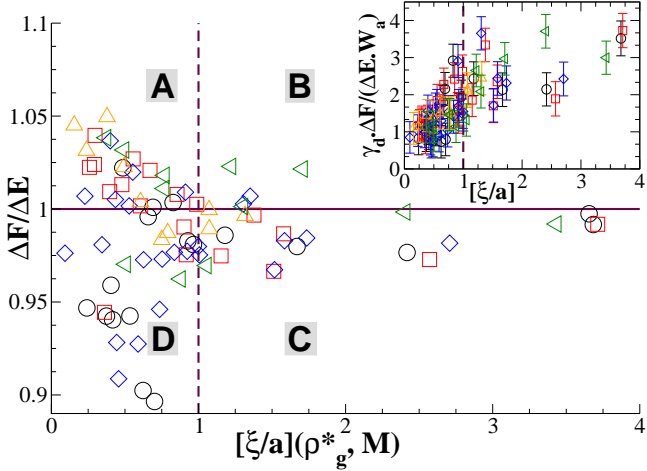

\rsfig{figure11.eps}
\caption{(Color online) Cross-over diagram between the free energy cost $\Delta E$ of deforming a solid surface and a liquid along the lines of Eq.~\eqref{DeltaE}, and the free energy cost $\Delta F$ of deforming an elastic soft surface from Eq.~\eqref{DeltaF}, experienced by a spherical droplet. The ratio $\Delta F/\Delta E$ is plotted as a function of the cross-over ratio $\xi/a$ from the Eq.~\eqref{xi}, which define four different wetting regimes $\{\textbf{A,B,C,D}\}$ as discussed in the text. \emph{Inset:} scaling plot which points out the relevant role of the adhesion energy $W_a$ for droplets larger than $\xi$, which experience a coupling between surface tension effects and brush elasticity.}
\label{fig11}
\end{figure}

In Fig.~\ref{fig11}, we have plotted the ratio $\Delta F/ \Delta E$ as a function of $\xi/a$, for data obtained from simulations. Such a ratio gives a complete map of trends for the wetting of droplets over either a hard, liquid, or soft surface, respectively described by $\Delta E_{(i)}$, $\Delta E_{(ii)}$, and $\Delta F$. Note that the soft and liquid surfaces, within this model, mainly differ by the elastic energy term of deforming the surface. The different wetting regimes are labeled as follow, {\bf A}: for the droplet, wetting a hard surface is better than a soft surface, while the equilibrium shape is driven by a surface tension/elasticity coupling, {\bf B}: the same wetting regime is favorable but surface tension dictates the equilibrium droplet shape, {\bf C}: idem for the shape but it is more favorable for the droplet to wet a soft surface, and {\bf D}: the same wetting regime might be preferred whereas the droplet shape is achieved in a coupling between surface tension/elasticity. 

First we note that droplets smaller than $\xi$ tend to be located in regime {\bf B} for larger brush grafting densities and in {\bf C} otherwise.  The shape of the smallest drops does not strongly differ between a hard or a soft substrate, the minimization of the surface tension dictates the behavior. A balance is achieved between surface and adhesion energies, leading to the Young-Dupré equation~\cite{Brochard02}. Second, one observes for droplets of size larger than $\xi$ that increasing $\rhored{g}$ mainly drives the systems from regime {\bf D} to {\bf A}: the higher $\rhored{g}$ is, the larger must be the compatibility $\eps{bd}/\epsilon$ in order to for the melt to penetrate into the brush and reach a more favorable "soft-wetting" state, which is mainly driven by the competition between the brush elasticity term in Eq.~\eqref{DeltaF} and the brush/liquid energy strength in the Eq.~\eqref{DeltaE}. Because one can expect the adhesion energy $W_a$ to be proportional to $\eps{bd}$~\cite{Brochard02,Kardar99}, one can confirm this trend by rescaling the ratio $\Delta F/\Delta E$ by $\surft{d}/W_a$, where the droplet surface tension $\surft{d}\equiv\surft{LV}$. Results are summarized in the inset of Fig.~\ref{fig1}, where, despite of the statistical error of the simulation data, the data are compatible with a scaling behavior. Thus, the behavior in regimes {\bf D} and {\bf A} is mainly governed by the adhesion energy.

\section{Concluding remarks}
\label{sec:Conclude}

Molecular dynamics simulations were performed in order to study the properties of the solid/liquid/vapor interfaces between viscous droplets and a deformable surface. First, particular attention has been paid to characterize the soft polymer brush surface. To this end, an implementation of an AFM-like numerical experiment has been developed, which has allowed us to extract the shear modulus of the brush as a function of the grafting density for the brush. The simulation data agree the ones theoretically predicted by Fredrickson \emph{et al.}~\cite{Fredrickson92} and the experimental work of Fujii \emph{et al.}~\cite{Fujii10}.

The wetting properties of the surface are determined by its softness and  the compatibility between the droplets and the surface. The former can be controlled by varying the grafting density of the brush, the latter is strongly influenced by the compatibility between brush and liquid. Density contours have shown a deformation of the wetting ridge near to the three-phase contact line. The strength of this lifting-up was found to increase with the compatibility between brush and liquid, as well as with the capacity of deformation of the surface. 

Finally, the wetting properties of spherical droplets that are in contact with the same kind of deformable surface have been also investigated. Different drop sizes and brush grafting densities were considered. We estimate the line tension and the concomitant critical droplet radius, above which line-tension effects may not be decisive. Then, minimizing an effective interface Hamiltonian with effective parameters that mimic  the thermodynamic properties of the simulated systems, it has been possible to capture the lifting-up of the contact line, and to  recover the line tension effect for droplets of sizes larger or smaller than the previously mentioned critical radius.

The interplay between the contact properties of the spherical drops and the deformable surface and its impact on the droplets shape deformation has then been addressed within a similar effective interface Hamiltonian. As the surface/droplet compliance can be tuned independently to the surface softness, and also because one can deal with different droplet sizes, it has been possible to cover a wide range of wetting and deformation behaviors. Two distinct asymptotic regimes have emerged. For droplets smaller than a critical size, the contact properties and its impact on the drop shapes are mainly dictated by the balance between the adhesion and surface energies. In this limit, the drops equilibrium properties are well characterized by the Young-Dupré formalism. In an other limit, for larger radii, the balance of surface tension and brush elasticity dictates the wetting behavior, and adhesion energies finally determine the equilibrium droplet shapes. 

The same kind of size effect has been shown to have a significant role on the dynamical properties of rolling droplets over hard surfaces~\cite{Servantie08}. It is then of interest to extend the current study to the steady state motion of drops over a soft surface, first to evaluate the contribution of the lifting-up effect to the dissipation of the moving contact line, and then to address droplet-size effects of the dissipation mechanisms. Further investigations are also required to fully characterize the slippage of the soft surfaces when complex fluids low past it.

\acknowledgments
The authors thank A.~Galuschko for useful comments. This work has been supported by the priority program \emph{Nano- and microfluidics} of the German Science Foundation (DFG) under grant Mu1674/3. Ample computer time at the computing centers JSC Jüllich, HLRN Hannover and Berlin, as well as the GWDG Göttingen are gratefully acknowledge. Part of the computing work was performed using the LAMMPS code~\cite{Plimpton95}, available under GPL at http://lammps.sandia.gov/.


\begin{thebibliography}{10}
\bibitem{Bocquet07} L.~Bocquet and J.-L.~Barrat, {\it Soft Matter} \textbf{3}, 685 (2007).
\bibitem{Stone} E. Lauga, M.P.~Brenner and H.A.~Stone, {\it Microfluidics: The no-slip boundary condition}, in {\it Handbook of Experimental Fluid Dynamics}, J. Foss, C. Tropea and A, Yarin (Hrsg), Springer, New York Chapter 15, cond-mat/0501557 (2008).
\bibitem{Leonforte11} F.~Leonforte, J.~Servantie, C.~Pastorino and M.~M\"uller, {\it J. Phys.: Condens. Matter}, \textbf{23}, 184105 (2011).
\bibitem{PJG} P. G. de Gennes, {\it Rev. Mod. Phys.} \textbf{57}, 827 (1985). 
\bibitem{BW91} C. Redon, F. Brochard-Wyart and F. Rondelez, {\it Phys. Rev. Lett.} \textbf{ 66}, 715 (1991).
\bibitem{Reiter92} G. Reiter, {\it Phys. Rev. Lett.} \textbf{68}, 75 (1992).
\bibitem{Karim} R. Xie, A. Karim, J. F. Douglas, C. C. Han and R. A. Weiss, {\it Phys. Rev. Lett.} \textbf{81}, 1251 (1998). 
\bibitem{Green03} P.F.~Green and V.~Ganesan, {\it Eur. Phys. J. E} \textbf{12}, 449 (2003).
\bibitem{Fetzer06} R.~Fetzer \emph{et al.}, {\it Europhys. Lett.} \textbf{75}, 638 (2006).
\bibitem{Fetzer05} R.~Fetzer \emph{et al.}, {\it Phys. Rev. Lett.} \textbf{95}, 127801 (2005).
\bibitem{Baumchen09} O.~Bäumchen, R.~Fetzer and K.~Jacobs, {\it Phys. Rev. Lett.} \textbf{103}, 247800 (2009).
\bibitem{Rev1} R.V. Craster, O.K. Matar, {\it Rev. Mod. Phys.} \textbf{81}, 1131 (2009).  
\bibitem{Young1805} T.~Young, {\it Philos. Trans. R. Soc. London} \textbf{95}, 65 (1805).
\bibitem{Navier1823} C.L.M.H.~Navier, {\it Mem. Acad. Roy. Sci. Inst. France} \textbf{6}, 389 (1823).
\bibitem{Shanahan94} M.E.R.~Shanahan and A.~Carré, {\it Langmuir} \textbf{10}, 1647 (1994).
\bibitem{Carre95} A.~Carré and M.E.R.~Shanahan, {\it Langmuir} \textbf{11}, 24 (1995).
\bibitem{Carre96} A.~Carré, J.-C.~Gastel, and M.E.R.~Shanahan, {\it Nature} \textbf{379}, 432 (1996).
\bibitem{Lester61} G.R.~Lester, {\it J. Coll. Int. Sci.} \textbf{16}, 315 (1961).
\bibitem{Das2011} S.~Das, A.~Marchand, B.~Andreotti, and J.H.~Snoeijer, {\it arXiv:1103.0782}.
\bibitem{Kumar04} S.~Kumar and O. K.~Matar, {\it J. Coll. Interf. Sci.}, \textbf{273}, 581 (2004). 
\bibitem{D1} Y.~Xu,	W.C.~Engl, E. R.~Jerison, K.J.~Wallenstein, C.~Hyland, L.A.~Wilen and E.R.~Dufresne, {\it Proc. Natl. Acad. Sci. U.S.A.} \textbf{107}, 14964 (2010).
\bibitem{D2} E.R.~Jerison,Y.~Xu, L.A.~Wilen and E.R.~Dufresne, {\it Phys. Rev. Lett.} \textbf{106}, 186103 (2011).
\bibitem{Carre01} A.~Carré and M.E.R.~Shanahan, {\it Langmuir} \textbf{17}, 2982 (2001).
\bibitem{Kremer90} K.~Kremer and G.S.~Grest, {\it J. Chem. Phys.} \textbf{92}, 5057 (1990).
\bibitem{DPD1} P. J. Hoogerbrugge and J. M. V. A. Koelman, {\it Europhys. Lett.} \textbf{19}, 155 (1992). 
\bibitem{DPD2} P. Warren and P. Espanol, {\it Europhys. Lett.} \textbf{30}, 191196 (1995 ). 
\bibitem{Rector94} Rector, Van Swol, Henderson, {\it Molecular Physics} \textbf{82}, 1009 (1994).
\bibitem{Kremer88} K.~Kremer, G.S.~Grest and I.~Carmesin, {\it Phys. Rev. Lett.} \textbf{61}, 566 (1988).
\bibitem{Everaers04} R.~Everaers \emph{et al.}, {\it Science} \textbf{203}, 823 (2004).
\bibitem{Pastorino07} C.~Pastorino, T.~Kreer, M.~Müller, and K.~Binder, {\it Phys. Rev. E} \textbf{76}, 026706 (2007).
\bibitem{Servantie08} J.~Servantie and M.~Müller, {\it J. Chem. Phys.} \textbf{128}, 014709 (2008).
\bibitem{Yong09} X.~Yong and L.T.~Zhang, {\it Langmuir} \textbf{25}, 5045 (2009).
\bibitem{MullerPlathe99} F.~Müller-Plathe, {\it Phys. Rev. E} \textbf{59}, 4894 (1999).
\bibitem{Tenney10} C.M.~Tenney and E.J.~Maginn, {\it J. Chem. Phys.} \textbf{132}, 014103 (2010).
\bibitem{Sen05} S.~Sen, S.K.~Kumar, and P.~Keblinski, {\it Macromolecules} \textbf{38}, 650 (2005).
\bibitem{Nijmeijer88} M.J.P.~Nijmeijer \emph{et al.}, {\it J. Chem. Phys.} \textbf{89}, 3789 (1988).
\bibitem{Orea03} P.~Orea, Y.~Duda, and J.~Alejandre, {\it J. Chem. Phys.} \textbf{118}, 5635 (2003).
\bibitem{Gretz1966} R.D.~Gretz, {\it J. Chem. Phys.} \textbf{45}, 3160 (1966).
\bibitem{MacDowell02} L.G.~MacDowell, M.~Müller, and K.~Binder, {\it Colloids Surf. A} \textbf{206}, 277 (2002).
\bibitem{Grest99} G.S.~Grest, {\it Adv. Polym. Sci.} \textbf{138}, 149 (1999).
\bibitem{Dimitrov08} D.I.~Dimitrov, A.I.~Milchev, and K.~Binder, {\it Macromol. Theory Simul.} \textbf{17}, 313 (2008).
\bibitem{Buff1965} F.P.~Buff, R.A.~Lovett, F.H.~Stillinger, {\it Phys. Rev. Lett.} \textbf{15}, 621 (1965).
\bibitem{Pastorino09} C.~Pastorino, K.~Binder and M.~Müller, {\it Macromolecules} \textbf{42}, 401 (2009).
\bibitem{Muller1996} M.~Müller and M.~Schick, {\it J. Chem. Phys.} \textbf{105}, 8885 (1996).
\bibitem{Utz08} M.~Utz and M.R.~Begley, {\it J. Mech. Phys. Solids} \textbf{56}, 801 (2008).
\bibitem{Fredrickson92} H.G.~Fredrickson, A.~Adjari, L.~Leibler, and J.-P.~Carton, {\it Macromolecules} \textbf{25}, 2882 (1992).
\bibitem{Williams93} D.R.M.~Williams, {\it Macromolecules} \textbf{26}, 5096 (1993).
\bibitem{Halperin10} A.~Halperin and E.B.~Zhulina, {\it Langmuir} \textbf{26}, 8933 (2010).
\bibitem{Akgun07} B.~Akgun, B.R.~Lee, H.~Kim, H.~Zhang, O.~Prucker, J.~Wang, J.~Rühe, and M.D.~Foster, {\it Macromolecules} \textbf{40}, 6361 (2007).
\bibitem{Long96} D.~Long, A.~Ajdari, and L.~Leibler, {\it Langmuir} \textbf{12}, 1675 (1996).
\bibitem{Fondecave98} R.~Fondecave and F.~Brochard-Wyart, {\it Macromolecules} \textbf{31}, 9305 (1998).
\bibitem{Fujii10} Y.~Fujii, Z.~Yang, A.~Clough, and O.K.C.~Tsui, {\it Macromolecules} \textbf{43}, 4310 (2010).
\bibitem{Dimitrov10} D.I.~Dimitrov, A.I.~Milchev, and K.~Binder, {\it Phys. Rev. E} \textbf{81}, 041603 (2010).
\bibitem{Saville77} G.~Saville, {\it J. Chem. Soc., Faraday Trans. 2} \textbf{73}, 1122 (1977).
\bibitem{Milchev01_1} A.I.~Milchev and K.~Binder, {\it J. Chem. Phys.} \textbf{114} 8610 (2001).
\bibitem{Milchev01_2} A.I.~Milchev and A.A.~Milchev, {\it Europhys. Lett.} \textbf{56}, 695 (2001).
\bibitem{Weijs11} J.H.~Weijs, A.~Marchand, B.~Andreotti, D.~Lohse, and J.H.~Snoeijer, {\it Phys. Fluid.} \textbf{23}, 022001 (2011).
\bibitem{DeConinck08} J.~De Coninck and T.D.~Blake, {\it Annu. Rev. Mater. Res.} \textbf{38}, 1 (2008).
\bibitem{Muller03} M.~Müller and L.G.~MacDowell, {\it J. Phys.: Condens. Matter} \textbf{15}, R609 (2003).
\bibitem{Muller2011} M.~Müller, Chapter in {\it Comprehensive Polymer Science}, K.~Matyjaszewski and M.~Möller (edts), Vol. {\bf 1} {\it Basic Concepts and Polymer Properties} edited by L.~Leibler and Al.~Khokhlov, Elsevier, Oxford, UK, 2011.
\bibitem{Muller2002} M.~Müller, {\it Comp. Phys. Comm.} \textbf{147}, 292 (2002).
\bibitem{Jerison2011} E.R.~Jerison, Y.~Xu, L.A.~Wilen, and E.R.~Dufresne, {\it Phys. Rev. Lett.} \textbf{106}, 186103 (2011).
\bibitem{Berg10} J.K.~Berg, C.M.~Weber, and H.~Riegler, {\it Phys. Rev. Lett.} \textbf{105}, 076103 (2010).
\bibitem{Brochard02} P.-G.~de Gennes, F.~Brochard-Wyart, D.~Quere, {\it Capillarity and Wetting Phenomena}, Springer, New-York, 2002.
\bibitem{Kardar99} M.~Kardar and R.~Golestanian, {\it Rev. Mod. Phys.} \textbf{71}, 1233 (1999).
\bibitem{Plimpton95} S.~Plimpton, {\it J Comp Phys} \textbf{117}, 1-19 (1995). 
\end{thebibliography}

\end{document}